# Demodulating Digital Holograms with Unknown Uniform Phase-shifts by Spiral Phase Transform

Pavel Trochtchanovitch

## Abstract

The Spiral Phase Transform (SPT) is a generalization of the Hilbert transform for 2D signals and, as such, can be used for AC signal demodulation. However, phase demodulation with the SPT is complicated by a multiplicative term that depends on the fringe directional map. We derived an analytical formula for the twofold directional map and applied the SPT for the blind reconstruction of phase-shifted digital holograms. Possible phase ambiguities in the unfolded directional map were resolved by satisfying the spatial uniformity condition of the phase shifts. The method was experimentally verified using on-axis and off-axis digital holograms of specularly reflecting subjects.

## 1. Introduction

A phase-shifted digital holography (PS-DH) is a versatile technique for wavefront sensing. It has a rich history and is deeply rooted in its predecessor phase-shifting interferometry, which appeared in 70s [1], and sometimes it's hard to draw a line between these 2 techniques. Unlike single-shot off-axis digital holography, PS-DH allows full-resolution wavefront sampling, which makes it attractive to many applications despite a limiting requirement of taking multiple exposures to reconstruct a single wavefront. But since the beginning of phase-shifted interferometry and up until most recently developed methods of PS-DH one common problem keeps re-emerging. For an accurate estimation of the wavefront the phase-shifts should be known precisely. In reality those phase-shifts may be altered by both instrumental imperfections (hysteresis and/or non-linearity of a phase shifter) and the environment (acoustic noise, mechanical vibrations, air flow and such). Thus, in practice the phase-shifts should be estimated from the same experimental dataset as the wavefront itself. Many methods had been proposed and refined to solve this problem. Most of them may be attributed to several broad categories, including: iterative methods [2, 5, 8, 9, 12, 13, 31], among which the most popular one is the "advanced iterative algorithm" (*AIA*, [8]), Lissajous ellipse fitting (*LEF*) [3, 4, 30, 33, 34, 39, 44], spatial averaging [14, 15, 17, 26, 27, 38, 46], inner products [22, 42], Singular Value Decomposition (*SVD*) [18, 19, 25, 32, 36, 40], optimization [10, 11, 16, 24, 49], interferogram comparison [28, 29, 37], and Kalman filter for locally quadratic phase [43].

Further developments had been made by combining some of those methods: SVD and least squared errors fitting [21], SVD and AIA [23], LEF and AIA [35], SVD and LEF [41], inner product and LEF [47]. A recent review of LEF methods can be found in [50].

But most of those methods require either some additional *a priori* information about the wavefront (constant intensity of reference wave [10, 12, 14, 16, 17, 49], slowly varying background and/or modulation components [18, 22, 46], locally quadratic phase [43]) or additional measurements of wave intensities ([15, 28, 29]). The methods that don't have such limitations either require many interferograms to be captured with different phase-shifts ([3, 4, 30, 33, 34, 39, 44]) or need a lot of data processing with many iterations (like AIA) before the right solution is found ([2, 5, 8, 9, 12, 13, 31]). Though it has been shown that AIA can be effectively vectorized for parallel processing on GPU with ~ 500x time speed-up [48], it still requires significant computational resources.

Among other methods an elegant approach was presented in [20], where the phase of a single interferogram was estimated with the help of Spiral Phase Transform (*SPT*), [7]. The directional map, which is required to establish the interferogram's phase from *SPT*, was found with the help of the second phase-shifted interferogram and a regularized optical flow calculation. This method is simple and fast, but, unfortunately, not always very accurate. Here we propose its modification that uses a different approach to establish the directional map and delivers more precise results.

The method proposed here doesn't require any *a priori* information about the reference and object waves, needs just 2 interferograms to estimate the phase and a scaled modulation amplitude (if the scaling factor has to be established, and/or a background term estimation is needed, a third interferogram with a different phase-shift is required). The only requirements for this method to be applicable are: either noticeable carrier frequency should present in the interferogram spectrum (the digital hologram has to be off-axis) or the phases of the object and the reference waves should be continuous functions of spatial coordinates, so that the fringe directional map is also continuous. These requirements are hardly limiting so the proposed method covers a broad range of practical applications.

The remainder of the article has the following stricture. Section 2 explains the theory of SPT demodulation in the context of its application to phase-shifted holograms. Section 3 provides methods for phase-shifts estimation from results of demodulation obtained in Section 2; Section 4 presents experimental results, and conclusions are made in Section 5; a list of references is provided in Section 6. The Matlab code illustrating the proposed method can be found at:
1) https://www.mathworks.com/matlabcentral/fileexchange/184172-uniform-phase-shifts-and-spt
2) https://github.com/pavelTrow/Uniform-Phase-Shifts-and-SPT

## 2. Demodulating 2D Signals with the Spiral Phase Transform (SPT)

To measure the wavefront in phase-shifted digital holography (PS-DH) *N* interferograms are captured with relative phase-shifts, which are created by introducing additional optical path difference between the reference and the object waves. The intensities of the interferograms are digitized by image sensor and *k*-th interferogram intensity can be expressed as:

$$I_{Hk} = I_B(x, y) + I_A(x, y) \cdot \cos(\phi(x, y) + \delta_k) \quad , \tag{2.1}$$

where $I_B(x, y) = I_R(x, y) + I_O(x, y)$ is the DC-term (also called "background"), $I_R(x,y)$ is the reference wave intensity; $I_O(x, y)$ is the object wave intensity;
$I_A(x, y)$ is the modulation amplitude, if the phase of the interference is stable during the recording, then $I_A(x, y) = 2 \cdot \sqrt{I_R \cdot I_O}$ ; $\phi(x, y) = \phi_O(x, y) - \phi_R(x, y)$ is the relative phase, a difference between phases of object and reference wave; $\delta_k$ is a phase-shift specific to the *k-th* interferogram.

The general task of digital hologram reconstruction is to demodulate captured interferograms, that is from a set of measured intensities $I_{Hk}(x, y)$ and *a priori* knowledge of $I_R(x, y)$ and $\phi_R(x, y)$ reconstruct object wave's phase and intensity $\phi_O(x, y)$ and $I_O(x, y)$ respectively. Here we will be solving an equivalent problem of finding phase, modulation amplitude and, possibly, background term from a set of interferograms (2.1).

For demodulation of those interferograms we need to suppress the DC-term first. Assuming that illumination is stable enough, so laser's power change between exposures is negligible this can be done by subtracting *k*-th interferogram intensity from the zeroth (the "reference") one:

$$\delta I_{Hk} \stackrel{\text{def}}{=} I_{H0} - I_{Hk} = 2 \cdot I_A(x, y) \cdot \sin\left(\frac{\delta_k - \delta_0}{2}\right) \cdot \sin\left(\phi(x, y) + \frac{\delta_k + \delta_0}{2}\right) \quad .$$

Without any loss of generality we can calculate the phase relative to the "reference" interferogram, so $\delta_0 = 0$, and we can rewrite the interferogram differences as follows:

$$\delta I_{Hk} = I_{Ak}(x, y) \cdot \sin(\phi_k(x, y)) \quad , \tag{2.2}$$

where $I_{Ak} = 2 \cdot I_A(x, y) \cdot \sin\left(\frac{\delta_k}{2}\right)$ , $\phi_k(x, y) = \phi(x, y) + \frac{\delta_k}{2}$ .

Here we are measuring $\delta_k$ modulo $2\pi$, so $0 < \delta_k < 2\pi$ , and $\sin(\delta_k / 2) > 0$ .

Our task now is to find amplitude $I_{Ak}(x,y)$ and phase $\phi_k(x,y)$ of the AC signal (2.2). In the 1D case this problem is solved by Hilbert transform. For 2D cases a good alternative is a Spiral Phase Transform (SPT) proposed for 2D signal demodulation tasks in [7]. The SPT is defined for AC signals with zero average (zero absolute value of 2D FFT transform at zero frequency) and can be calculated as:

$$\text{SPT}\{.\} \stackrel{\text{def}}{=} \text{FT}^{-1}\left\{\left(\frac{\omega_x + j \cdot \omega_y}{\sqrt{\omega_x^2 + \omega_y^2}}\right) \cdot \text{FT}\{.\}\right\} \quad ,$$

where FT{.} stands for 2D Fourier transform; the phase factor has a singularity at zero frequency, and that's why we require that the FT of our AC signal is zero there.

When applying this transform to AC signal of the interferogram (2.2) we will get [7]:

$$I_{Ck} = \text{SPT}\{\delta I_{Hk}\} = -j \cdot \exp(j \cdot \eta(x, y)) \cdot I_{Ak}(x, y) \cdot \cos(\phi_k(x, y)) \quad , \tag{2.3}$$

where ***j*** designates an imaginary unit, $\eta(x,y)$ – is a directional map; at every pixel it defines a direction perpendicular to the local fringe direction. Please, note that this map is expected to be the same for all interferograms since it doesn't depend on the phase-shifts.

From (2.2), (2.3) we can find the scaled amplitude modulation term $I_{Ak}(x,y)$ since $\cos^2(\phi_k(x, y)) + \sin^2(\phi_k(x, y)) = 1$ :

$$I_{Ak} = \sqrt{(\delta I_{Hk})^2 + |I_{Ck}|^2} \quad , \tag{2.4}$$

where | z | stand for absolute value of z.

For phase estimation we need to determine the directional map $\eta(x,y)$ first. In [20] it was proposed to obtain this map by capturing an additional interferogram with small positive phase-shift and calculating the optical flow between the two interferograms, which provide an approximate result with accuracy acceptable in many cases. Here we propose a more rigorous solution.

For that we will find sine and cosine quadratures of the *twofold directional map*. We can do that thanks to the fact that $\cos^2(\phi_k)$ is non negative, so having it as multiplication terms in quadratures won't affect the angle estimation (except pixels, where $\cos(\phi_k)$ is zero). From (2.3) we have:

$$\Re(I_{Sk})\cdot\Im(I_{Sk}) = -I_{Ak}^2\cdot\sin(\eta)\cdot\cos(\eta)\cdot\cos^2(\phi_k) = -\frac{I_{Ak}^2}{2}\sin(2\eta)\cdot\cos^2(\phi_k) \quad ,$$

where $\Im(z)$ , $\Re(z)$ are imaginary and real parts of *z* respectively. From which, since $\cos^2(\phi) = 1 - \sin^2(\phi)$ , we finally get:

$$\sin(2\eta) = -\frac{2\cdot\Re(I_{Sk})\cdot\Im(I_{Sk})}{I_{Ak}^2 - (\delta I_{Hk})^2} \tag{2.5}$$

Also, from (2.3): $\Im(I_{Sk})^2 = I_{Ak}^2\cdot\cos^2(\eta)\cdot\cos^2(\phi_k)$ , so

$$\cos(2\eta) = 2\cdot\cos^2(\eta) - 1 = \frac{2\cdot\Im(I_{Sk})^2 - I_{Ak}^2 + (\delta I_{Hk})^2}{I_{Ak}^2 - (\delta I_{Hk})^2} \tag{2.6}$$

If we define *twofold direction map* modulo 2π as:

$$\theta \overset{\text{def}}{=} \operatorname{mod}(2\cdot\eta + \pi, 2\pi) - \pi \quad , \tag{2.7}$$

we can find it from (2.5) and (2.6) as: follows

$$\theta = \operatorname{atan2}\left(-2\cdot\Re(I_{Sk})\cdot\Im(I_{Sk}), \quad 2\cdot\Im(I_{Sk})^2 - I_{Ak}^2 + (\delta I_{Hk})^2\right) \quad , \tag{2.8}$$

where atan2(y, x) – is the angle measure between the positive x-axis and the ray from the origin to the point (x, y) in the Cartesian plane. For a complex number z: $\arg(z) = \operatorname{atan2}(\Im(z), \Re(z))$ .

The expression (2.8) is not defined if $\cos(\phi_k) = 0$ . Thus, in the presence of noise we may expect huge errors in $\theta$-map estimation (2.8) for pixels with $\cos(\phi_k) \approx 0$ . This is a small subset of hologram pixels, and we can find $\theta$-map values for them by extrapolation from the neighbor pixels. Practically, for example, an individual twofold directional map (2.8) can be filtered with 2D *spatial* median filter.

However, having captured multiple interferograms we can suppress the noise in the twofold directional map by *temporal* median filter as follows. For every *k*-th interferogram difference $\delta I_{Hk}(x,y)$ (2.2) we calculate the corresponding twofold directional map $\theta_k(x,y)$ by (2.8). Then taking an odd number $Q$ of $\theta$-maps we form a 3D array of twofold directional map by concatenating the estimated maps in the 3rd dimension:

$$\Theta(x,y,k)=[\theta_1(x,y);\theta_2(x,y);...;\theta_k(x,y);...;\theta_Q(x,y)] \quad .$$

The filtered twofold directional map is found then for every pixel as a median value along that dimension:

$$\theta_F(x,y)=\underset{k}{\text{median}}\left(\Theta(x,y,k)\right) \quad . \tag{2.9}$$

Since we require the phase-shift values to be different, the noisy areas of $\theta_k$ – maps with $\cos(\phi_k)\approx 0$ will be different for different *k*. Thus, the noise should be effectively removed by (2.9).

It is worth noting that in general case we can not unfold the direction map $\eta(x,y)$ by simply dividing the *twofold map* $\theta(x,y)$ from (2.8) by 2. Since $\theta$ is defined modulo 2π we may potentially miss a $\pm\pi$ term at every pixel. But this problem can be circumvented in two practically important cases:

- The twofold direction map $\theta(x,y)$ has limited range of angles;
- The twofold direction map $\theta(x,y)$ is a continuous function of spatial coordinates.

In the former case we can estimate the direction map as the half of the twofold one:

$$\eta(x,y)=\frac{\theta_F(x,y)}{2} \quad . \tag{2.10a}$$

The estimation (2.10a) may have $\pm\pi$ error, but this would be the same error for *all* hologram pixels, and its net effect will be reversal of the global sign of hologram's phase $\phi_k(x,y)$. This sign is typically known *a priori* so its inversion will be captured and corrected.

In the latter case, when directional map is a continuous function of spatial coordinates, the $\pi$ – ambiguity can be resolved by unwrapping the twofold directional map $\theta(x,y)$, and then dividing the unwrapped map by 2:

$$\eta(x,y)=\frac{1}{2}\cdot\text{UNW}\left(\theta_F(x,y)\right) \quad , \tag{2.10b}$$

where UNW(.) is an unwrapping operator that removes 2π discontinuities from the measured wrapped $\theta_F(x,y)$ phase map. There are many techniques for phase unwrapping developed so far. In this work we are using the "discreet cosine transform" (*DCT*) method to unwrap the phase, which is based on the numerical solution of the Poisson's equation for phase [6].

After $\eta(x,y)$ is found, from (2.3) we can find the cosine quadrature for the hologram phase:

$$J_{Ck}=j\cdot I_{Ck}(x,y)\cdot\exp(-j\cdot\eta)=I_{Ak}(x,y)\cdot\cos\left(\phi_k(x,y)\right) \quad . \tag{2.11}$$

And finally from (2.2) and (2.10) we find the phase (modulo 2π):

$$\phi_k=\text{atan}\,2\left(\delta I_{Hk},\, J_{Ck}\right) \quad , \tag{2.12}$$

which solves the hologram demodulation task.

It’s worth noting that during demodulation procedures we never masked hologram spectrum, as typically done in conventional off-axis digital holography, and used all the pixel, so the final resolution of the amplitude $I_{Ak}(x,y)$ and phase $\phi_k(x,y)$, obtained from (2.4) and (2.12) respectively, is the same as of the input interferograms.

After demodulating interferogram difference (2.2) according to formulas (2.4), (2.12) we will get estimations of a scaled modulation term and a shifted phase:

$$I_{Ak}=2\cdot I_A(x,y)\cdot\sin\left(\frac{\delta_k}{2}\right) \quad , \tag{2.13a}$$

$$\phi_k(x,y)=\phi(x,y)+\frac{\delta_k}{2} \quad . \tag{2.13b}$$

For many practical tasks this information is enough as far as the hologram reconstruction problem is concerned. Please note that to find those values we needed just two interferograms.

The problem with the 2-interferograms approach is that any errors in the directional map estimation (2.10 a, b) will be directly propagating to the phase estimation (2.13b).

With the 3rd interferogram we will be able to determine the phase-shifts, and then use standard techniques to estimate the actual modulation term $I_A(x,y)$, the DC-term $I_B(x,y)$, and the phase. By this approach the errors on the directional map estimation may be corrected, thanks to spatial uniformity of the phase-shifts, as demonstrated in sections 3, 4. The next section provides the details on this refinement process.

# 3. Estimating Phase-shifts, DC-term, and Modulation Amplitude

## *3.1 First Estimation of the Phase-shifts*

If we need to estimate the phase-shifts $\delta_k$ the demodulation process (2.3), (2.4), (2.8-2.12) should be repeated for interferogram differences with another “reference interferogram”. Here we described this method for $N = 3$ interferograms. The generalization for a larger number of interferograms is straightforward.

First, using the “zeroth” interferogram as the reference one from (2.3), (2.4), (2.8-2.12) we get the following quantities:

$$I_{A1}=2\cdot I_A(x,y)\cdot\sin\left(\frac{\delta_1}{2}\right) \quad , \qquad I_{A2}=2\cdot I_A(x,y)\cdot\sin\left(\frac{\delta_2}{2}\right) \quad , \tag{3.1.1a}$$

$$\phi_1(x,y)=\phi(x,y)+\frac{\delta_1}{2} \quad , \qquad \phi_2(x,y)=\phi(x,y)+\frac{\delta_2}{2} \quad . \tag{3.1.1b}$$

Then we will calculate the difference between the first and the second interferograms:

$$\delta I_{21}\stackrel{\text{def}}{=}I_{H1}-I_{H2}=2\cdot I_A(x,y)\cdot\sin\left(\frac{\delta_2-\delta_1}{2}\right)\cdot\sin\left(\phi(x,y)+\frac{\delta_2+\delta_1}{2}\right) \quad . \tag{3.1.2}$$

Here, again, we are assuming that the phase-shifts are defined modulo 2π, and also require the condition $\delta_2 > \delta_1$ to be satisfied, which can be guaranteed by the proper experiment design. After repeating the SPT demodulation procedure for (3.1.2) we will obtain:

$$I_{A\,21} = 2 \cdot I_A(x,y) \cdot \sin\left(\frac{\delta_2 - \delta_1}{2}\right) \quad , \tag{3.1.3a}$$

$$\phi_{21}(x,y) = \phi(x,y) + \frac{\delta_2 + \delta_1}{2} \quad . \tag{3.1.3b}$$

From (3.1.1b), (3.1.3b) we can find the phase relative to the first interferogram:

$$\phi = \phi_2 + \phi_1 - \phi_{21} \quad . \tag{3.1.4}$$

After the phase is found we can find “half-phase-shift maps” as corresponding differences:

$$\Delta_k(x,y) \stackrel{\text{def}}{=} \text{mod}\left(\phi_k(x,y) - \phi(x,y), 2\pi\right) \quad . \tag{3.1.5}$$

If the phase-shifts are uniform we should get almost constant values of $\Delta_k$ with some amount of random noise. If some pixels have $\Delta_k(x,y)$ noticeably different from its median value, it is a clear indication that those areas have errors in directional map estimation $\eta(x,y)$ (2.10).

Assuming that the percentage of pixels with those errors are small we can get a good first estimation of the phase-shifts as:

$$\delta_k^{(0)} = 2 \cdot \underset{x,y}{\text{median}}\left(\Delta_k(x,y)\right) \quad . \tag{3.1.6}$$

As the next step we will find problematic areas in the directional map, apply correction, and perform the next iteration of phase-shift estimation, based on corrected directional maps, as described in the next section.

### *3.2 Directional Map Correction and Phase-shifts Refinement*

As it was mentioned previously, we expect our phase-shift to be uniform over interferogram areas, so large variations in $\Delta_k(x,y)$ are clear indications that there are errors in the directional map. To find those areas we will set thresholds:

$$\epsilon_k \stackrel{\text{def}}{=} |\delta_k^{(0)} - \pi| \quad , \tag{3.2.1}$$

where the first estimations of phase-shifts $\delta^{(0)}_k$ are defined by (3.1.6).

Then we will define individual error indicator functions as:

$$\sigma_k(x,y) = \begin{cases} 1, & if \quad |\Delta_k(x,y) - \delta_k^{(0)}| > \epsilon_k \\ 0, & otherwise \end{cases} \quad .$$

The net error indicator function is found as their multiplication:

$$\sigma \stackrel{\text{def}}{=} \prod_k \sigma_k(x,y) \tag{3.2.2}$$

Since our twofold directional map $\theta(x,y)$ is estimated correctly, though modulo 2π, the errors in the unfolded directional map may only be equal to π. So the corrected directional map can be found as:

$$\eta_{CORR}(x,y)=\eta(x,y)+\pi\cdot\sigma(x,y) \quad , \tag{3.2.3}$$

After the directional map is corrected the phases can be re-calculated with (2.11), (2.12), (3.1.3b), (3.1.4), and half-phase-shift maps (3.1.5) refined. Finally, the updated estimation for the phase-shifts is found with (3.1.6):

$$\delta_k=2\cdot\underset{x,y}{\text{median}}\left(\Delta_k^{CORR}(x,y)\right) \quad . \tag{3.2.4}$$

With the known phase-shifts (3.2.4) it is straightforward to find the DC-term and AC modulation amplitudes, as described in the next section.

## *3.3 Estimating Phase, DC-term and AC modulation with Known Phase-shifts.*

The problem of estimating interferograms' phase while minimizing the phase-noise was solved in [45]. In this work authors derived expressions for the optimal coefficients to reconstruct the phase and modulation amplitude from a set of noisy interferograms (2.1), when the phase-shifts are known.

The expressions obtained in [45] for hologram quadratures are the following:

$$\begin{aligned} I_{\sin}&=\frac{1}{N^2}\cdot\sum_{k=0}^{N-1}\sum_{l=0}^{N-1}\sum_{m=0}^{N-1} I_{Hm}\cdot(\cos\delta_k-\cos\delta_l)\cdot \mathrm{D}(k,l,m) \quad , \\ I_{\cos}&=\frac{1}{N^2}\cdot\sum_{k=0}^{N-1}\sum_{l=0}^{N-1}\sum_{m=0}^{N-1} I_{Hm}\cdot(\sin\delta_k-\sin\delta_l)\cdot \mathrm{D}(k,l,m) \quad , \end{aligned} \tag{3.3.1}$$

where $\mathrm{D}(k,l,m)=\cos\delta_k\cdot\sin\delta_l+\cos\delta_m\cdot\sin\delta_k+\cos\delta_l\cdot\sin\delta_m$ , $\delta_0 = 0$,
$I_{\cos}\overset{\text{def}}{=}I_A(x,y)\cdot\cos(\phi(x,y))$ , $I_{\sin}\overset{\text{def}}{=}I_A(x,y)\cdot\sin(\phi(x,y))$ .

Thus, the optimal modulation amplitude can be obtained from quadratures (3.3.1) as:

$$I_{AOPT}=\sqrt{I_{\cos}^2+I_{\sin}^2} \quad . \tag{3.3.2}$$

An optimal estimation for the phase would be:

$$\phi_{opt}=\text{atan}\,2(I_{\sin},I_{\cos}) \quad . \tag{3.3.3}$$

For the DC-term we can get $N$ estimations from:

$$I_B^{(k)}(x,y)=I_{Hk}-I_{AOPT}(x,y)\cdot\cos(\phi_{opt}(x,y)+\delta_k) \quad , \tag{3.3.4}$$

and then find the mean value:

$$I_B(x,y)=\frac{1}{N}\sum_{k=0}^{N-1} I_B^{(k)}(x,y) \quad . \tag{3.3.5}$$

The equations (3.3.1) – (3.3.5) solve the hologram demodulation task. The practical application of the proposed method to experimental data is presented in the next section.

## 4. Experimental Results

To validate the proposed method we applied it to ROIs of phase-shifted digital holograms obtained in Michelson-type setup from a PCB with specular reflecting metallic IC. The subject was focused on the image sensor. Four interferograms had been captured with intended relative phase-shifts of $\pi k/2$ with k = 0, 1, 2, 3. However, due to imperfections of the phase shifter and impacts from the environment the phase-shifts were actually different, so here they are assumed to be unknown.

The first experiment described in section 4.1 illustrates application of the proposed method to an on-axis hologram with continuous directional map, so (2.10b) can be used to estimate this map.

The second example in section (4.2) illustrates the application of the proposed method to the off-axis case with dominating direction in the directional map, which in this case can be calculated from (2.10a).

### *4.1 Continuous Directional Map Example*

To validate the proposed method for on-axis digital holography we selected parts of interferograms with zero carrier frequency (Fig. 4.1.1). Those parts correspond to a specular reflecting surface of a metallic IC, so both the interferograms' phase and the directional map are expected to be continuous functions of spatial coordinates. As a result the twofold directional map can be unwrapped and equation (2.10b) can be used to estimate the unfolded directional map.

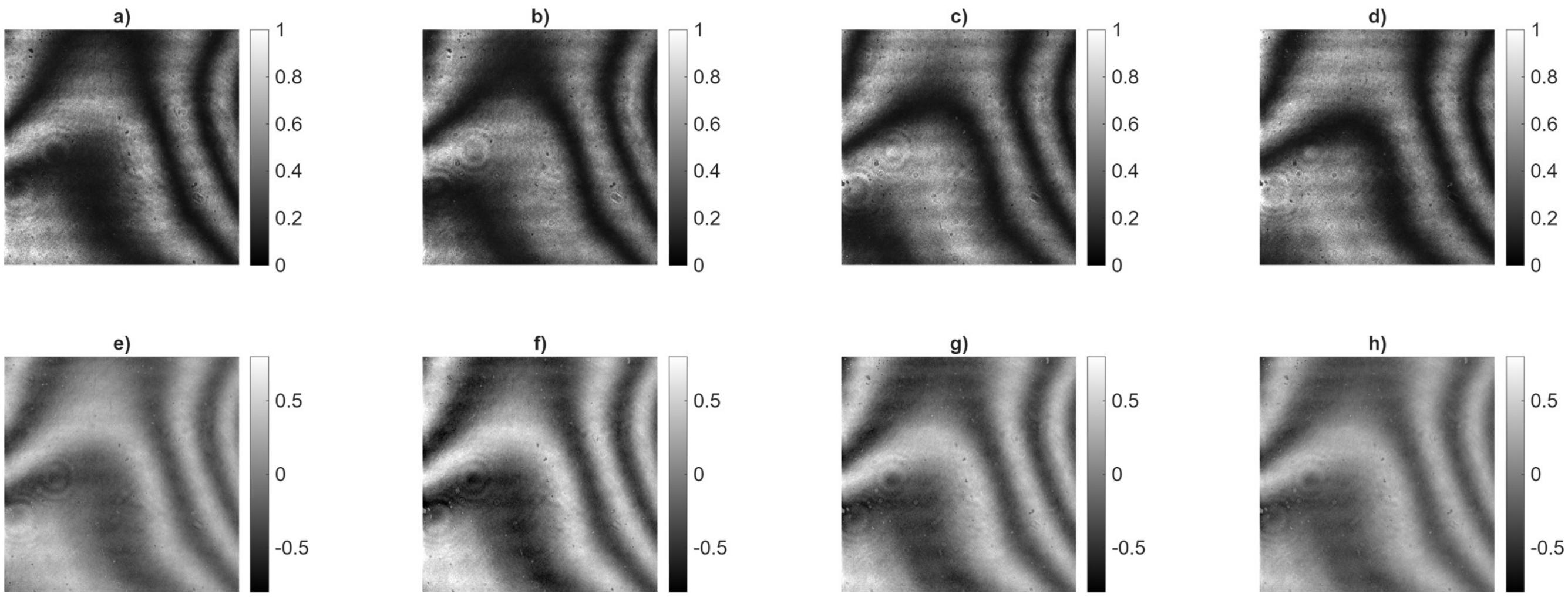

**Fig. 4.1.1. Input Interferograms and Their Differences; a), b), c), d) – $I_{Hk}(x, y)$, k = 0, 1, 2, 3; e), f), g) - $\delta I_{Hk}(x, y)$, k = 1, 2, 3; h) - $\delta I_{21}$.**

The initial estimation of the twofold directional map is presented in Fig. 4.1.2. Subplots (a), (b), (c) show individual $\theta$-maps, which have noisy "lines" corresponding to pixels with $\cos(\phi_k) \approx 0$, as expected. Subplot (d) demonstrates that the temporal median filter (2.9) effectively suppresses this noise. Subplot (e) shows the unwrapped twofold directional map, calculated by DCT method [6]. As it will soon become apparent this map has unwrapping errors in the upper left part (blue area in Fig. 4.1.2(e)).

After unwrapping the filtered twofold directional map, we found the unfolded directional map from (2.10b), cosine quadratures from (2.11), and phases from (2.12). The process was repeated for the difference $\delta I_{21}$ (Fig. 4.1.1(h)) with a different "reference" interferogram as well. Then the "base" phase relative to the "zeroth" interferogram was found from (3.1.4) and the half-phase-shift maps were estimated from (3.1.5).

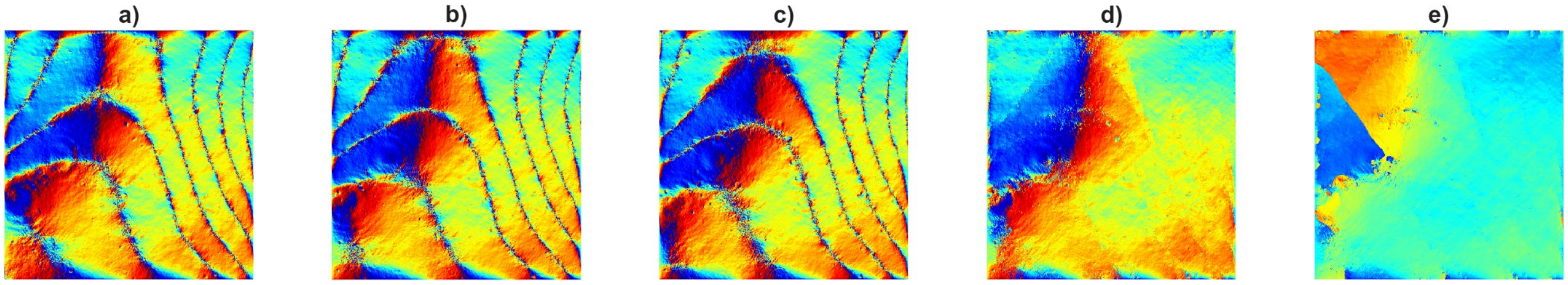

**Fig. 4.1.2. Twofold Directional Map Estimations;**
**a), b), c) – $\theta$–maps obtained for interferogram differences (2.2) for k = 1, 2, 3;**
**d) – temporally filtered $\theta$–map (2.9); e) – unwrapped filtered $\theta$–map.**

The first estimation of the half-phase-shift maps is presented in Fig. 4.1.3 along with their histograms. It is clear that those maps have errors caused by unwrapping errors of the twofold directional map (Fig. 4.1.2(e)), which can be seen as blue patches on the left. Those error patches can be removed first from the twofold directional map, and then from the half-phase-shift maps, as illustrated by Figs. 4.1.4, 4.1.5.

Fig. 4.1.4(a) shows the unfolded directional map, calculated by (2.10b), subplot (b) shows error indicator function (3.2.2), where errors are marked with black pixels. Finally, subplot (c) shows a corrected directional map (3.2.3). The half-phase-shift maps recalculated with the corrected directional map are presented in Fig. 4.1.5.

By comparing Figs. 4.1.3 and 4.1.5 we can see that half-phase-shift maps obtained with the corrected directional map are much more uniform, so much more accurate values of the phase-shifts can be obtained with (3.2.4).

Hologram reconstruction result for the refined phase-shift values is presented in Fig. 4.1.6, where the (a) subplot shows DC-term $I_B(x, y)$ (3.3.5); the (b) subplot shows AC modulation amplitude $I_{AOPT}(x, y)$ (3.3.2); and subplots (c, d, e) illustrate phase estimations for different stages of the algorithm.

Fig. 4.1.6(c) shows the first estimation of the phase (3.1.1b) before the directional map correction. The π - errors of the directional map are clearly visible as discontinuities in that phase estimation. Fig. 4.1.6(d) shows the same estimation with the corrected directional map; the phase is continuous now. Finally, Fig. 4.1.6(e) presents the "averaged" phase (3.3.3) reconstructed with the suppressed phase noise.

By comparing Fig. 4.1.6 (d) and (e) we can clearly see the noise suppression effect of the optimized quadratures (3.3.1), (3.3.3) that were proposed in [45].

It is worth noting that obtained phase-shift maps provide vital information for quantitative phase imaging applications. Indeed, the variance in those maps allows us to estimate the accuracy of the measured phase.

In our next experiment we applied a proposed method to an off-axis digital hologram that had a discontinuity in the phase.

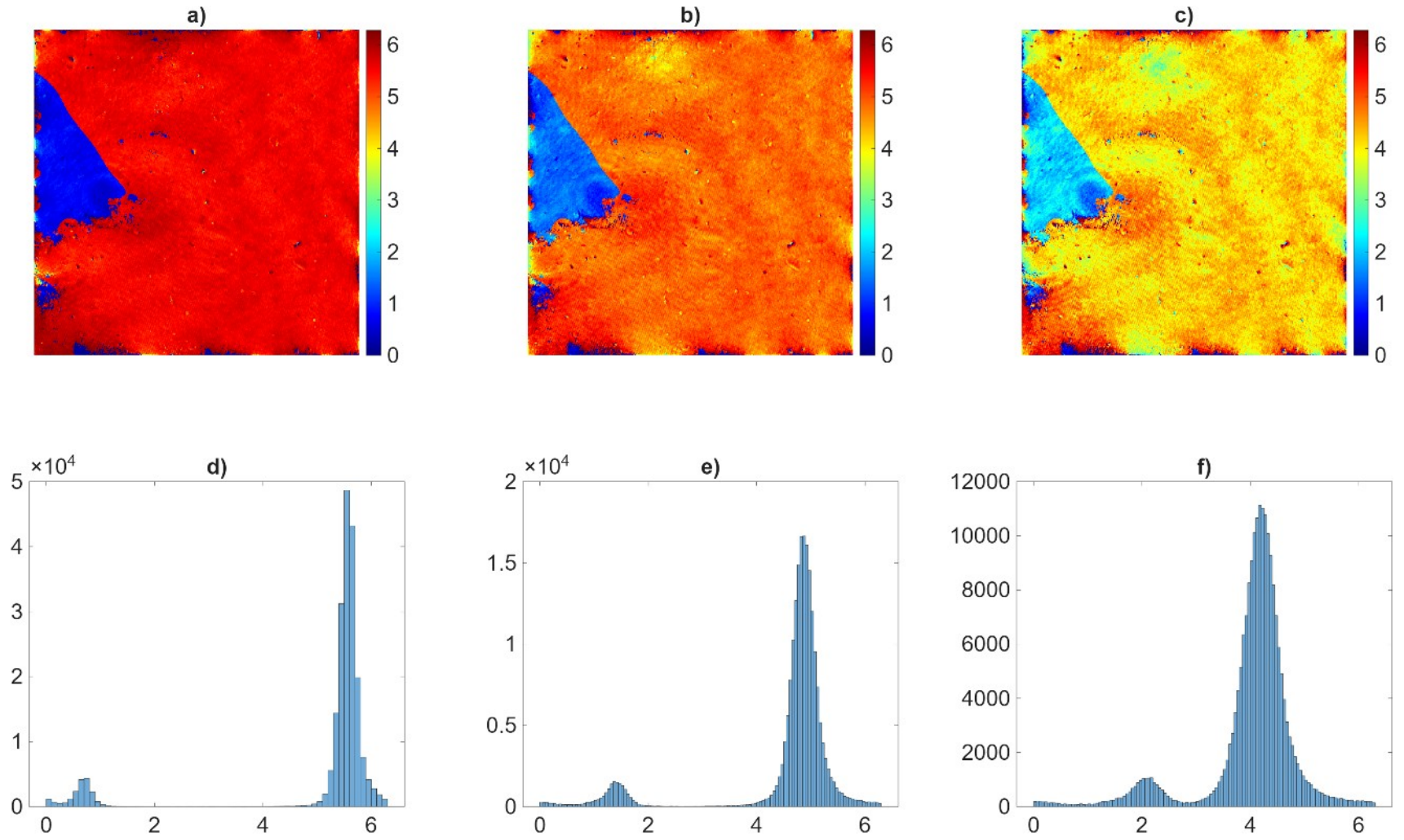

**Fig. 4.1.3. First Estimation of the Half-Phase-Shifts; a), b), c) – $\Delta_k(x,y)$, k = 1, 2, 3; d), e), f) – histograms of $\Delta_k(x,y)$, k = 1, 2, 3.**

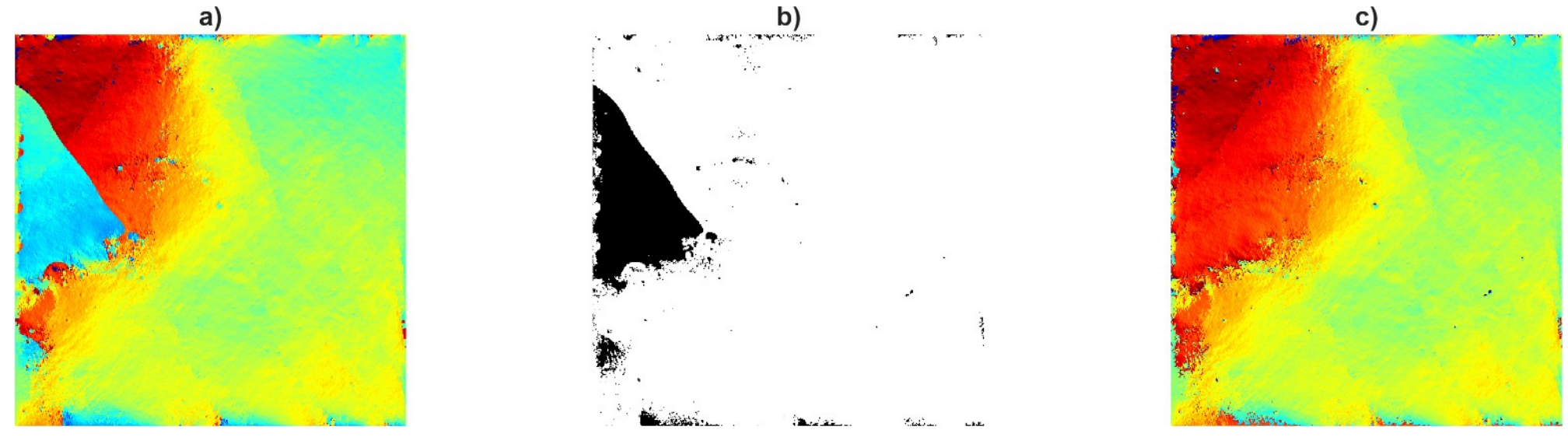

**Fig. 4.1.4. Directional Map Correction;**
**a) – First Estimation of Directional Map (2.10b); b) – Error Indicator (3.2.2);**
**c) – Corrected Directional Map (3.2.3).**

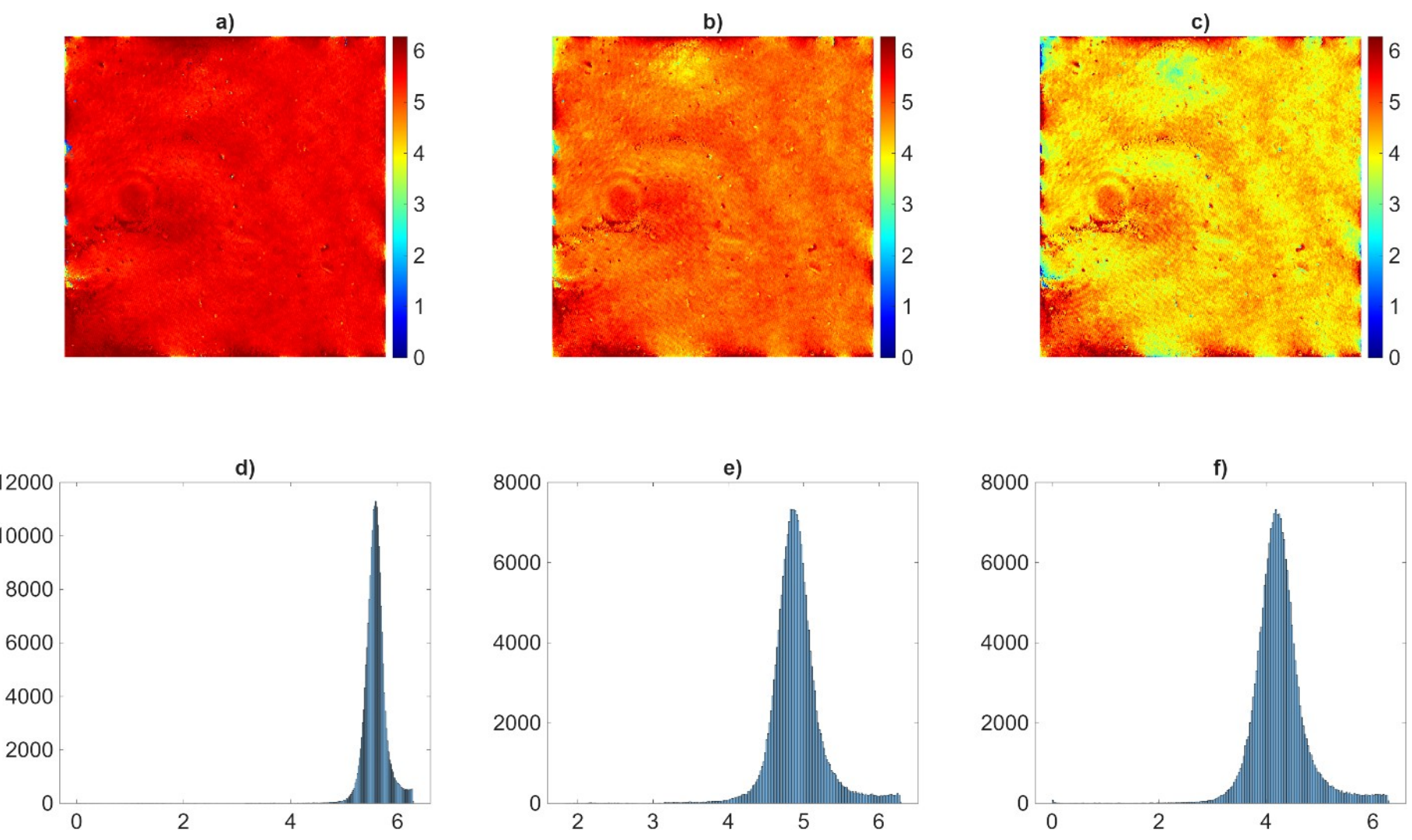

**Fig. 4.1.5. Corrected Half-Phase-Shifts; $\Delta^{CORR}{}_k(x,y)$, k = 1, 2, 3 (*top row*), and Their Histograms (*bottom row*).**

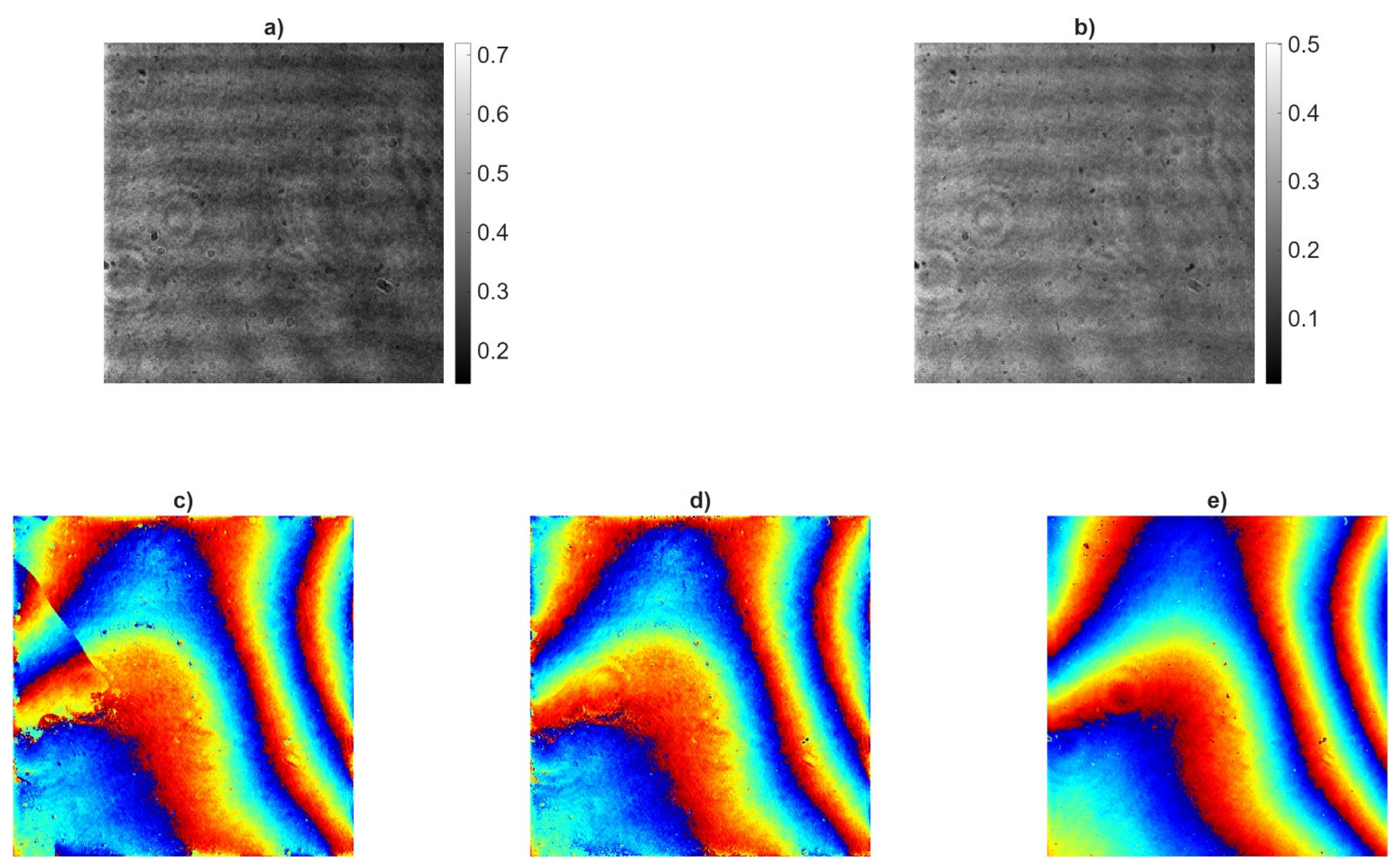

**Fig. 4.1.6. Digital Hologram Demodulation Results; a) – DC- term, b) – Modulation amplitude; c) First Estimation for Phase (3.1.4); d) Phase Estimation with Corrected Directional Map; e) Optimized Phase Estimation (3.3.3).**

## *4.2 Directional Map with Dominating Direction Example*

For our next example we consider a part of a partially off-axis digital hologram with the most of it having dominating fringe direction, so we can use (2.10a) to find the directional map from the twofold one. Fig. 4.2.1 presents the input interferograms (a, b, c, d) and their differences (e, f, g, h).

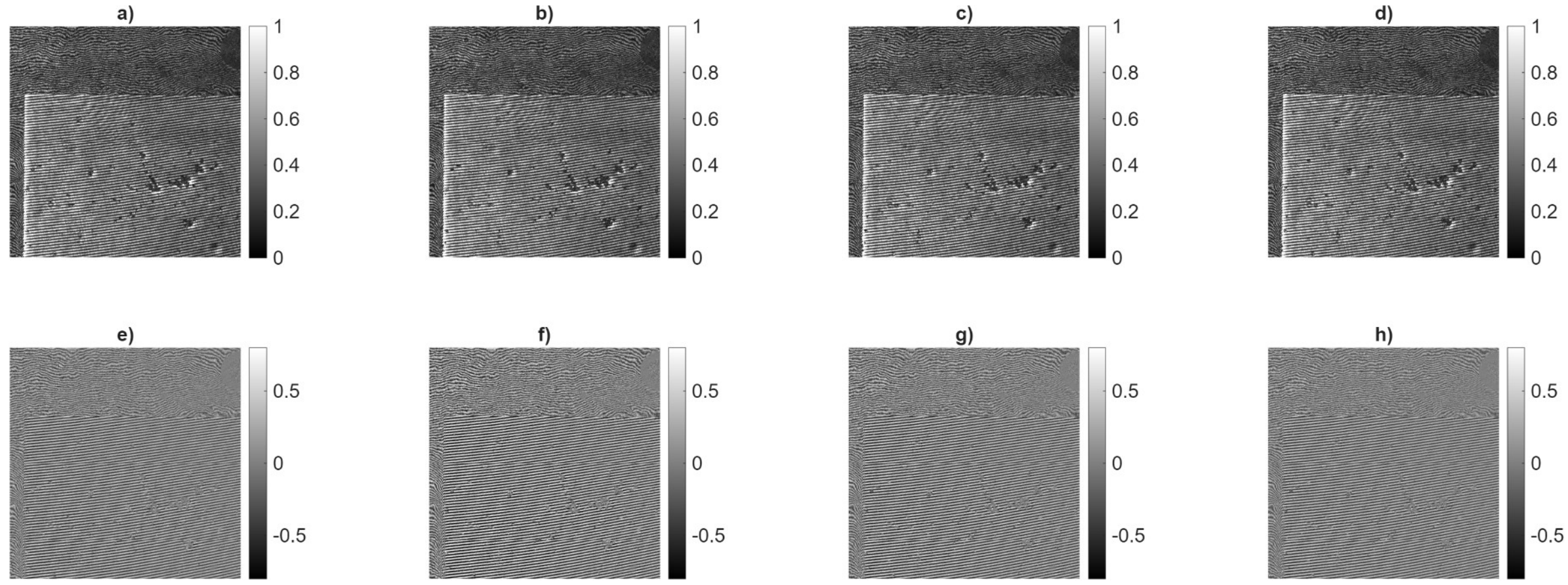

**Fig. 4.2.1. Input Interferograms and Their Differences; a), b), c), d) – $I_{Hk}(x, y)$, k = 0, 1, 2, 3; e), f), g) - $\delta I_{Hk}(x, y)$, k = 1, 2, 3;  h) - $\delta I_{21}$.**

The twofold directional maps calculated from (2.8) are presented in Fig. 4.2.2. The first 3 images (a, b, c) show the individual $\theta$-maps for differences $\delta I_{Hk}(x, y)$ with k = 1, 2, 3. Fig. 4.2.3(d) presents a temporally filtered twofold directional map, and its histogram is shown in Fig. 4.2.3(e).

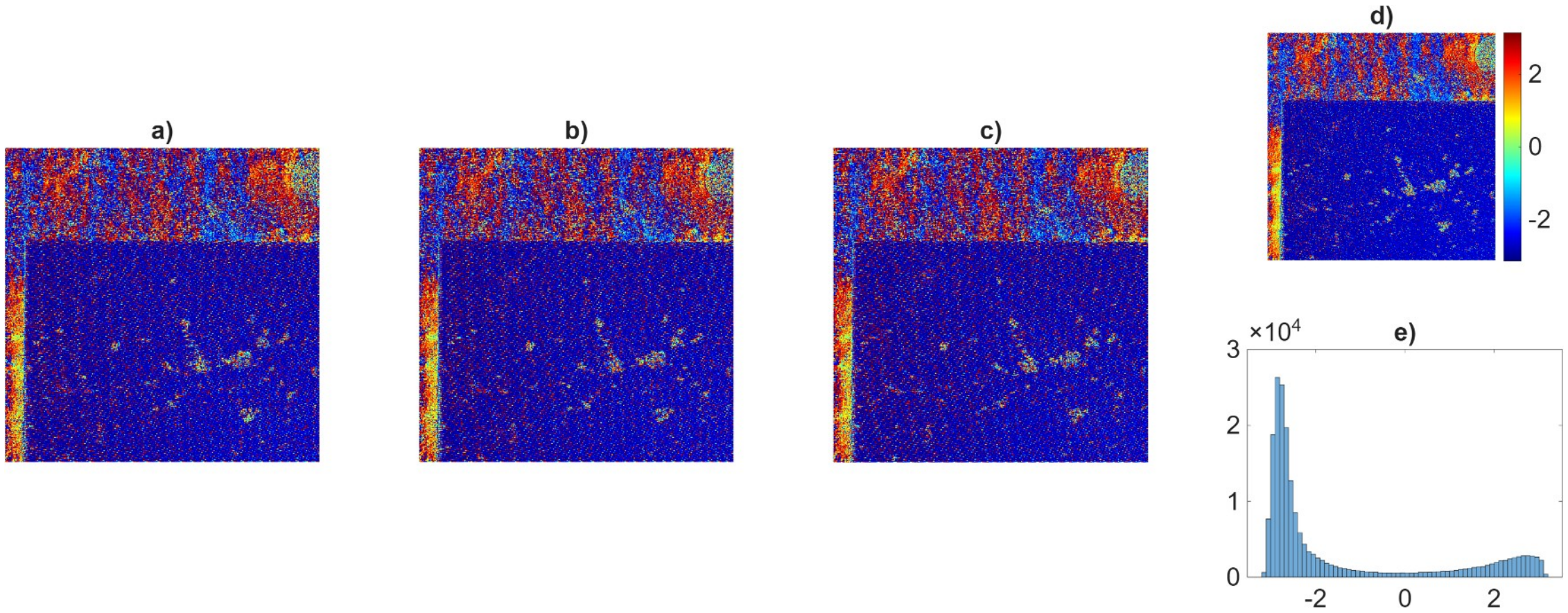

**Fig. 4.2.2. Twofold Directional Map Estimations;**
**a), b), c) – $\theta$–maps obtained for interferogram differences (2.2) for k = 1, 2, 3;**
**d) – temporally filtered $\theta$–map (2.9); e) – histogram of filtered $\theta$–map.**

From Fig. 4.2.2 we see that individual $\theta$-maps (a, b, c) have noisy lines that correspond to areas with $\cos(\phi_k) \approx 0$. However, the temporal median filter (2.9) effectively suppresses this noise (Fig. 4.2.3, (d)). The histogram of the filtered $\theta$-map (Fig. 4.2.3(e)) confirms that the map has a dominating direction, so we can use equation (2.10a) to find it.

Similarly to the previous section, Figs. 4.2.3, 4.2.4, and 4.2.5 present initial estimation of the half-phase-shift maps, correction of the directional map, and refined half-phase-shift maps respectively. By comparing Figs 4.2.3 and 4.2.5 we can see by how much the uniformity of the half-phase-shift maps are being improved by the directional map correction (3.2.3).

The refined phase-shifts estimated from maps of Fig. 4.2.5 by (3.2.4) were used to find hologram components: DC-term, AC modulation, and phase; as shown in Fig. 4.2.6. Similarly to Fig. 4.1.6, this figure shows hologram's phase at different stages of its reconstruction. For clarity the same constant ramp is removed from all phase estimations. The initial phase estimation is shown in Fig. 4.2.6(c); as can be seen from this figure, the areas with errors in directional map have very large noise (those areas are marked by black pixels in the error indicator map in Fig. 4.2.4(b)). However, for the phase estimation for the corrected directional map (Fig. 4.2.5(d)) this noise is removed and we can infer two "bumpy" surfaces: a PCB in the background and IC in the front. The averaged phase estimation from the optimized quadratures (3.3.1), which is shown in Fig. 4.2.6(e) has even less phase noise.

It is worth noting that though directional map errors directly affect the estimation of hologram phase, the estimation of its AC modulation amplitude (2.4) doesn't depend on the directional map, and its scaled version (3.1.1a) can be found from just 2 interferograms. To illustrate this property Fig. 4.2.7 shows AC modulation amplitude estimations (a, b, c) re-scaled by dividing them by $2 \cdot \sin(\delta_k/2)$ , along with optimal amplitude estimation (3.3.2), presented in Fig. 4.2.7(d). To quantitatively measure the similarity of those estimations we calculated normalized cross-correlation of the individual estimations Fig. 3.2.7(a, b, c) with the optimal one in Fig. 3.2.7(d); the normalized cross-correlation was found to be: 0.9927, 0.9934, and 0.9925 for estimations (a, b, and c) respectively, which proves high accuracy of amplitude estimation despite the errors in the directional map.

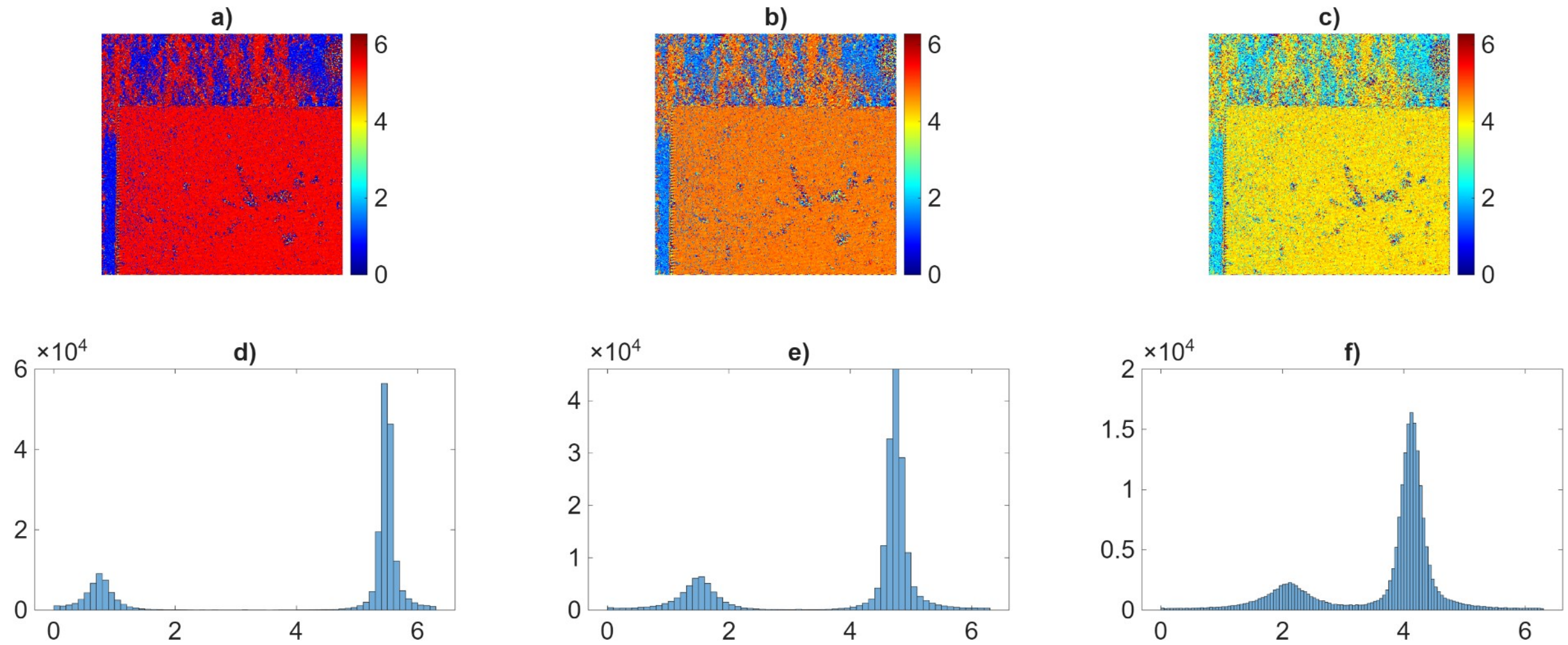

**Fig. 4.2.3. First Estimation of the Half-Phase-Shift Maps; a), b), c) – $\Delta_k(x,y)$, k = 1, 2, 3; d), e), f) – histograms of $\Delta_k(x,y)$, k = 1, 2, 3.**

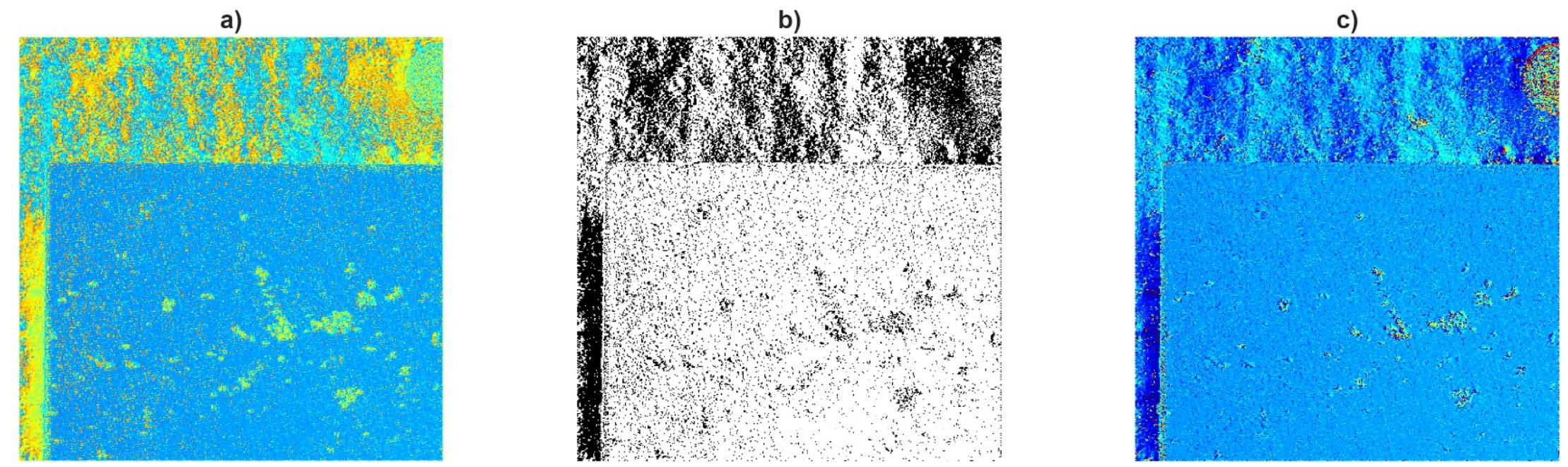

**Fig. 4.2.4. Directional Map Correction;**
**a) – First Estimation of Directional Map (2.10b); b) – Error Indicator (3.2.2);**
**c) – Corrected Directional Map (3.2.3).**

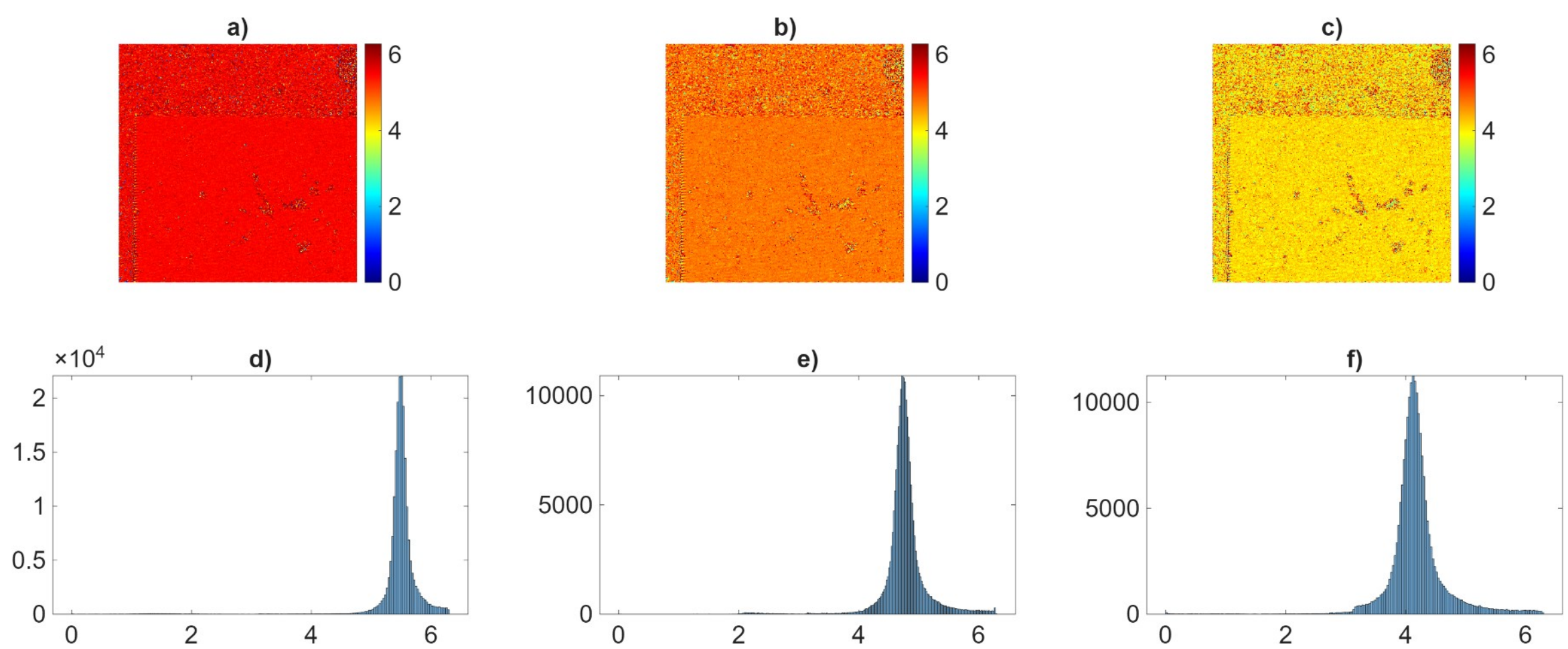

**Fig. 4.2.5. Corrected Half-Phase-Shift Maps; $\Delta^{CORR}_k(x,y)$, k = 1, 2, 3 (*top row*), and Their Histograms (*bottom row*).**

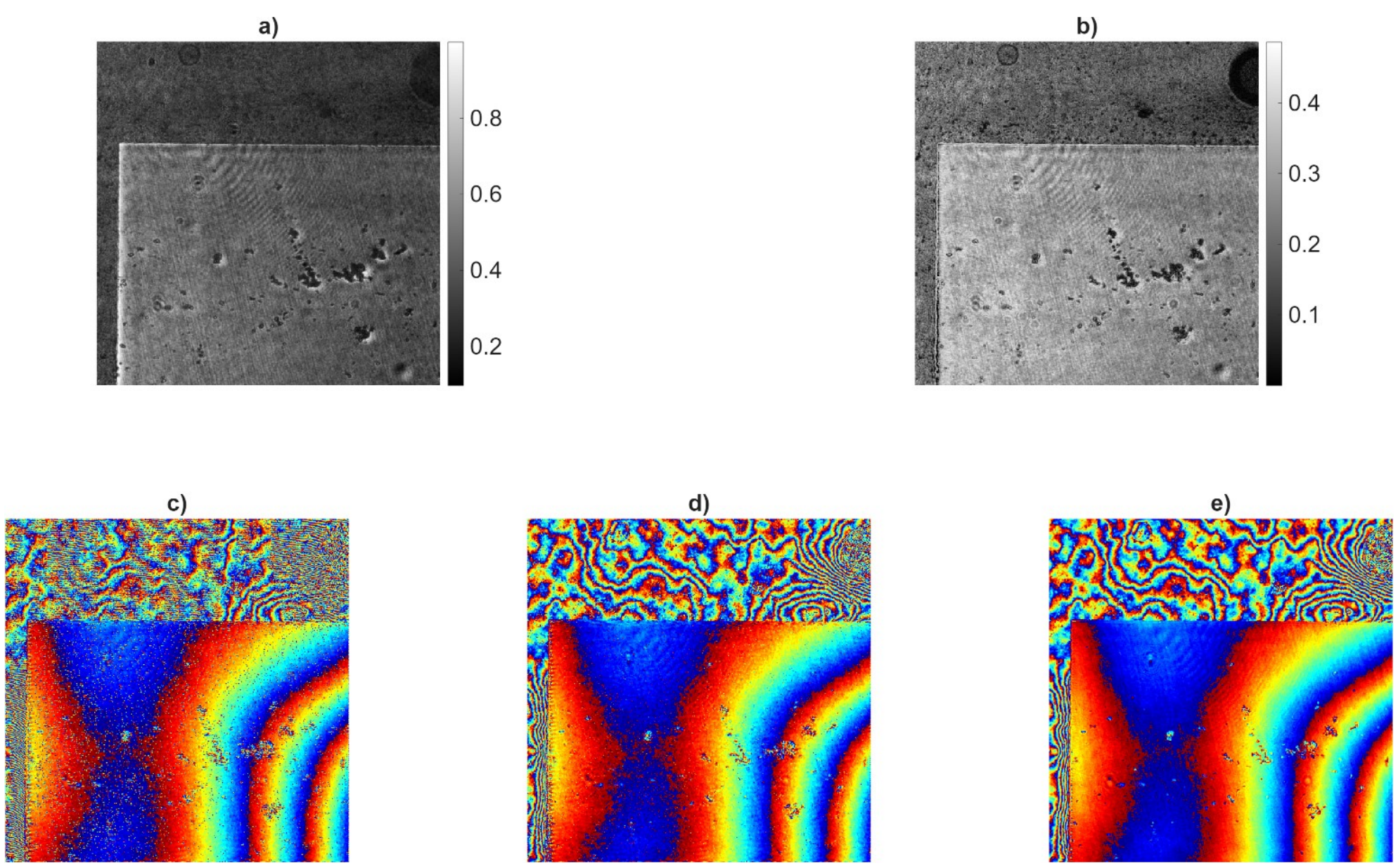

**Fig. 4.2.6. Digital Hologram Demodulation Results; a) – DC- term, b) – Modulation amplitude; c) First Estimation for Phase (3.1.4); d) Phase Estimation with Corrected Directional Map; e) Optimized Phase Estimation (3.3.3); same constant phase ramp is removed from the phase estimations for clarity.**

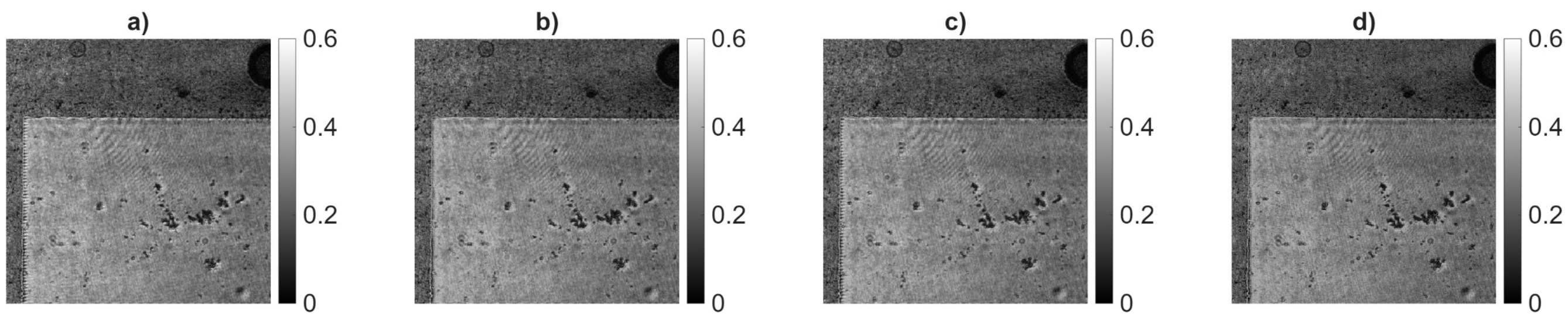

**Fig. 4.2.7. Modulation Amplitude Estimations;**
**a), b), c) – individual estimations (2.4) for k = 1, 2, 3 divided by** $2 \cdot \sin(\delta_k / 2)$ **;**
**d) – optimal estimation (3.3.2).**

## 5. Conclusions

The proposed method for blind demodulation of phase-shifted digital holograms is versatile and robust. It is better suited for off-axis (or partially off-axis) holograms, which have a directional map with dominating direction set by their carrier frequency, so that the directional map can be obtained from the twofold one by a simple division (2.10a). However, for on-axis holograms the method can also be applied, provided the phase of the interferograms is a continuous function of spatial coordinates. Thanks to the fact that full-resolution phase estimations can be obtained from SPT without any prior knowledge of the phase-shifts, the method is robust and allows to diagnose and correct errors in the directional maps, check consistency of modulation amplitudes in every interferogram pair, and predict accuracy of quantitative phase imaging by evaluating variance in the phase-shift maps.

The method can be further optimized for specific use cases. For example, for slowly varying intensities of reference and object waves, the DC term can be estimated by other methods reported previously (like polynomial fitting in [24], or Hilbert-Huang transform in [39]). In this case just one interferogram would be enough to get a full-resolution of the phase and amplitude of the object wave.

## Acknowledgments

The author would like to thank Dr. Dmitry Labukhin for reviewing the original manuscript and providing very valuable suggestions on its improvement.